\documentclass[fleqn]{article}
\usepackage{xspace,epsfig}
\begin{document}
\newcommand{\keywords}{coated conductors, critical current, superconducting tape}
\newcommand{\PACS}{74.76.Bz, 74.72.Bk, 85.25.Kx}
\newcommand{\shorttitle}{S.~Leitenmeier et al., Coated conductors
containing big grains} 
\newcommand{\unit}[2]{#1\,#2}
\newcommand{\DEG}[1]{\unit{#1}{\degree}}
\newcommand{\GB}{grain boundary}
\newcommand{\GBA}{\GB\ area}
\newcommand{\GBAs}{\GB\ areas}
\newcommand{\GBs}{grain boundaries}
\newcommand{\CC}{critical current}
\newcommand{\CCs}{critical currents}
\newcommand{\CCD}{\CC\ density}
\newcommand{\CCDs}{\CC\ densities}
\newcommand{\AR}{aspect ratio}
\newcommand{\ARs}{aspect ratios}
\newcommand{\AAR}{average \AR}
\newcommand{\AARs}{average \ARs}
\newcommand{\GN}{grain network}
\newcommand{\GNs}{\GN s}
\newcommand{\tenAcm}[1]{$10^{#1} \mathrm{A/{cm^{2}}}$}
\newcommand{\Acm}[2]{${#1}\cdot10^{#2} \mathrm{A/{cm^{2}}}$}
\newcommand{\Jcgrain}{$J_{\mathrm{c}}^{\mathrm{grain}}$}
\newcommand{\critical}[1]{$#1_{\mathrm{c}}$\xspace}
\newcommand{\Tc}{\critical{T}}
\newcommand{\Ic}{\critical{I}}
\newcommand{\Jc}{\critical{J}}
\newcommand{\scm}{cm$^2$\xspace}
\newcommand{\etal}{{\sl et\,al.}\xspace}
\newcommand{\eg}{{\sl e.\,g.}\xspace}
\newcommand{\ie}{{\sl i.\,e.}\xspace}
\newcommand{\YBCO}{YBa$_2$Cu$_3$O$_{7-\delta}$\xspace}
\newcommand{\nbco}{NdBa$_2$Cu$_3$O$_{7-\delta}$\xspace}
\newcommand{\ycbco}{Y$_{1-x}$Ca$_x$Ba$_2$Cu$_3$O$_{7-\delta}$\xspace}
\newcommand{\sto}{SrTiO$_3$\xspace}
\newcommand{\CeO}{CeO$_2$\xspace}
\newcommand{\MgB}{MgB$_2$\xspace}
\newcommand{\degree}{$^{\circ}$\xspace}

\title{Coated conductors
containing grains with big aspect ratios}
\author{\normalfont S.~Leitenmeier, H.~Bielefeldt, G.~Hammerl,
A.~Schmehl,\\
C.~W.~Schneider, and J.~Mannhart}
\newcommand{\address}
  {Experimentalphysik VI\\
   Center for Electronic Correlations and Magnetism\\
   Institute of Physics, Augsburg University, 86135 Augsburg, Germany}
\newcommand{\email}{}
\date{}
\maketitle
\begin{abstract}
It is shown that the critical currents of high-\Tc\ superconducting tapes fabricated by the coated
conductor technologies are enhanced considerably if grain arrangements with large effective grain
boundary areas are used. Increasing the \AR s of the grains reduces the deleterious effects of the
grain boundaries. A practical road to competitive high-\Tc cables is proposed.
\end{abstract}

\section*{}

\begin{figure}[tbp]
\centering \epsfig{file=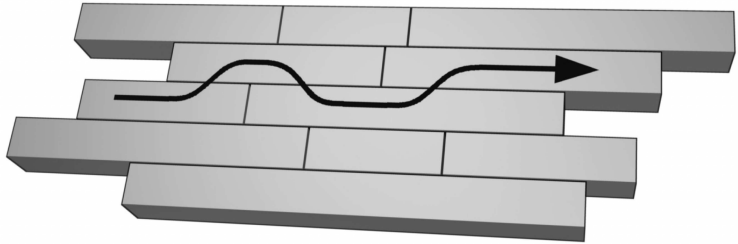,width=10cm} \caption{ Sketch of a
brick-wall like arrangement of grains in a coated conductor. The
arrow illustrates a possible path for the supercurrent. The
current traverses grain boundaries with large areas.}
\label{Fig:3DKorn}
\end{figure}

Cables that are superconducting at \unit{77}{K} require the use of
polycrystalline high-\Tc-superconductors \cite{bednorz} with large
\CCs. For materials with sufficient pinning, such as \YBCO
\cite{wu}, three strategies have been found to achieve this goal.
The first is to enhance the \textit{critical current density} of
the \GBs. This can be done by aligning the grains along all three
major axes to within few degrees \cite{Dimos}. This approach is
based on the fact that the \GB\ \CCD\ is an exponential function
of the misorientation angle \cite{Ivanov}, dropping by three to
four orders of magnitude as the misorientation angle is increased
from \DEG{0} to \DEG{45}. Second, for a given misorientation
angle, the \GB\ \CC\ density is enhanced by appropriate doping
\cite{Hammerl,Schmehl}. The third strategy consists in maximizing
the effective \textit{grain boundary area} by optimizing the
arrangement and the shape of the grains, as illustrated in Fig.
\ref{Fig:3DKorn}. In the simplest case, this can be achieved by
utilizing grains with large \ARs\
\cite{MannhartTsuei,MannhartSpringerBuch}.

The most promising candidates for economically competitive cables are tapes fabricated by coated
conductor technologies, such as ion beam-assisted deposition (IBAD) \cite{IBAD}, rolling-assisted
biaxially-textured substrates (RABiTS) \cite{RABiTS1,RABiTS2}, and inclined-substrate deposition
(ISD) \cite{ISD1,ISD2}. For practical applications, the coated conductor technologies are superior
to the competing powder in tube technology which is based on Bi-based high-\Tc superconductors
embedded in silver tubes, because the material costs of coated conductors are decisively smaller
and the ReBa$_2$Cu$_3$O$_{7-\delta}$\xspace superconductors, where Re is Y or a rare earth, offer
the potential of operation at 77 K in large magnetic fields. The best coated conductors fabricated
at present support \CCDs\ exceeding $10^{6} \mathrm{A/{cm^{2}}}$ over meter lengths. Taking the
substrate thickness and thereby the whole cross-section of the tape into account, these current
densities correspond to engineering \CCDs\ of several \tenAcm{4}. These values are achieved by
aligning the grains along all axes with a spread of misorientation angles smaller than \DEG{10}.
Because the corresponding grain alignment processes are slow and costly, strategies are urgently
sought to enhance the \CCD\ of such tapes for a given misorientation spread. The solution of this
problem would provide the key to commercially viable large scale applications of high-\Tc
superconductors.

Applying the concepts conceived in Ref.~\cite{MannhartTsuei} to coated conductors, we suggest to
enhance their \CCs\ \Ic\ by using grains with large \ARs\ to optimize the effective \GBA\ (see
Fig.~\ref{Fig:3DKorn}). According to the model calculations described below, an increase of the
average grain \AR\ causes a monotonous, strong increase of \Ic, as well as a reduction of the
sensitivity of \Ic\ to the average \GB\ angle.

The calculations performed to analyze the \CC\ of a given \GB\ network were based on a modified
version of the algorithm developed by Holzapfel {\it et al.}, as described in detail in
Ref.~\cite{Holzapfel}. This algorithm has been designed to analyze the \CCs\ of \GB\ networks in
coated conductors. The procedure considers two-dimensional grain networks characterized by a given
spread of grain orientations. In such a network the algorithm searches for the cross-section that
limits the \CC, and then calculates the \CC\ of this cross-section. Because for the grain
misorientations of interest the \GBs\ do not act as Josephson junctions, and all phase effects are
negligible. As shown in Ref. \cite{Holzapfel}, the results of such calculations agree well with
transport measurements of \Ic. For our work, the algorithm was optimized for speed, so that several
thousand networks, each containing $10^5 - 10^6$ grains, could be calculated on a personal
computer. This optimization was achieved by accelerating the search for the limiting cross-section,
storing information gained in the individual search routines.

\begin{figure}
\centering \epsfig{file=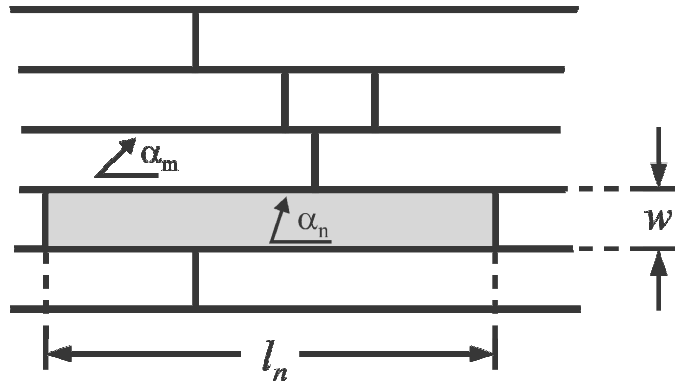,width=10cm} \caption{Illustration
of the grain arrangement considered in the calculations.}
\label{Fig:ModelGrain}
\end{figure}

\begin{figure}
\centering\hfill\epsfig{file=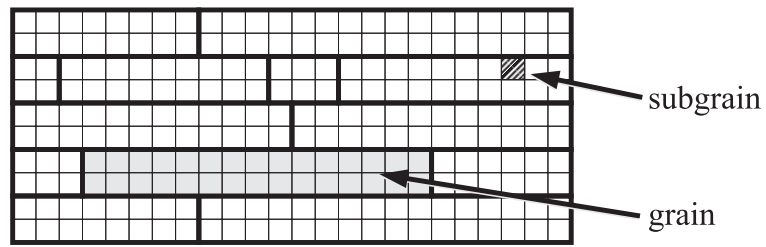,width=10cm} \caption{Sketch
of the grain and subgrain arrangement in a small array consisting
of 24\,*\,10 subgrains. The thick lines represent the grain
boundaries. The thin lines show the boundaries between the
subgrains.} \label{Fig:ModelSubGrain}
\end{figure}

To analyze the effect of the enhancement of the effective \GBA\ on \Ic, two-dimensional arrays of
$N$ grains arranged in brickwall-type structures were considered (see Fig.~\ref{Fig:ModelGrain} and
Fig.~\ref{Fig:ModelSubGrain}). In these structures to each grain an in-plane orientation $\alpha_i$
was assigned (see Fig.~\ref{Fig:ModelGrain}). Whereas the same width $w$ was selected for all
grains, the length $l_i$ of each grain was randomly chosen, following a Gaussian distribution with
a full width at half maximum (FWHM) of $l/5$ centered around the average length $l$, clipped to
zero below $w/2$ and above 250 $w$. The \AAR\ of the grains is then given by $N^{-1}\sum l_i/w$. As
the \CC\ of the network is also affected by the intragranular \CCD \Jcgrain, each grain was split
into two rows of square subgrains (see Fig.~\ref{Fig:ModelSubGrain}) and the \CCD\ of the
boundaries between the subgrains was set to equal \Jcgrain.

The angles $\alpha_i$ were also randomly chosen, again following a Gaussian distribution, this time
centered at \unit{0}{\degree} with a FWHM-spread $\sigma$. Misorientation angles $\alpha_i$
exceeding \DEG{45} were clipped.

The grain arrangements and the Gaussian distributions were selected to provide clear and simple
rules for the design of the model systems. These systems are presented as first examples for
practical conductors with more complicated designs. As will be obvious, the conclusions of our work
are not affected by the particular choice of the model systems used for the calculations.

In case self field effects are negligible, the \CC\ of the \GB\ between two grains, $n$ and $m$, is
well approximated by the product of the boundary area and its \CCD\ $J_{\mathrm{c}}^{n,m}$. To
calculate the \CCD, an exponential dependence on the misorientation angle $|\alpha_n-\alpha_m|$ was
used \cite{Ivanov}

\begin{equation}
J_{\mathrm{c}}^{n,m}=J_{\mathrm{c}}^{\mathrm{grain}}\cdot \exp
(-\frac{|\alpha_{n}-\alpha_{m}|}{\beta}).
\end{equation}

\begin{figure}
\centering \epsfig{file=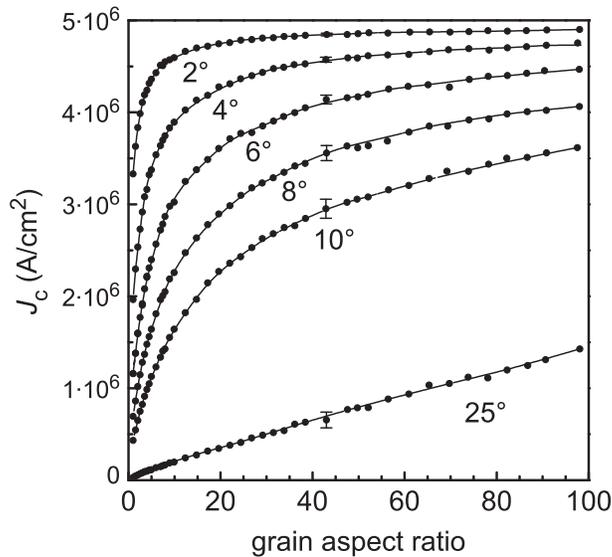,width=8cm}\hspace*{5mm}
\vspace{-5mm} \caption{Plot of the \CCD\ \Jc\ as a function of the
average grain \AR, calculated for various grain alignment spreads
$\sigma$. The data have been obtained in calculations on \GNs\
consisting of 1000\,*\,200 subgrains. Each datapoint reflects an
average of at least 20 calculations. The lines are guides to the
eye.} \label{fig3}
\end{figure}

Here \Jcgrain\ was set to \Acm{5}{6} and $\beta$ to \unit{5.3}{\degree}, values typical for coated
conductors operated at \unit{77}{K}. With this, the influence of the grain \AR\ on the \CC\ of the
network was analyzed. For a series of \AARs\ and grain orientation distributions the \CCDs\ of
networks 200 subgrains wide and 1,000 subgrains long were calculated. Depending on the grains \AAR\
these networks consisted of about 500 to 50,000 grains. For each set of parameters the \CCDs\ of at
least 20 different networks have been determined and the resulting \CCDs\ averaged. Fig.~\ref{fig3}
shows the resulting \CCDs. The error bars display the standard deviation of the calculated currents
as determined from the averaging process.

As shown by the figure, the \CCD\ of the network rises strongly and monotonously with increasing
\AR\ of the grains. The gradient is largest at small \ARs, which is attributed to the relatively
small current densities of these networks, which are barely affected by \Jcgrain. For small
misorientation angles and large \ARs\ the current densities saturate at \Jcgrain. The calculated
\Jc\ enhancements are impressive. For example, for grain misorientations resulting from an
alignment spread of \DEG{10}, the current densities are enhanced from \Acm{4.4}{5} to \Acm{1.6}{6}
if the \AAR\ is increased from 1 to 10.

\begin{figure}
\centering \epsfig{file=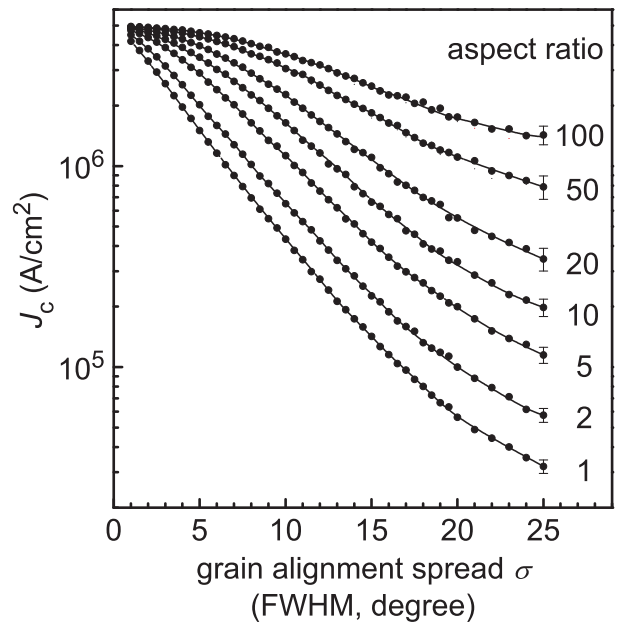,width=8cm}\hspace*{5mm}
\vspace{-2mm} \caption{Plot of the \CCD\ \Jc\ as a function of the
grain alignment spread $\sigma$, calculated for various average
grain \ARs. The data have been obtained in calculations on \GNs\
consisting of 1000\,*\,200 subgrains. Each datapoint reflects an
average of at least 20 calculations. The lines are guides to the
eye.} \label{fig4}
\end{figure}

Further, the dependence of the \CC\ on the spread $\sigma$ of the grain orientations was analyzed,
considering various \AARs. The results of these calculations are shown in Fig.~\ref{fig4}. As
expected, for small angular spreads the critical current densities of the networks equal the
intragrain critical current density, independent of the \ARs. Surprisingly, in networks consisting
of grains with large aspect ratios, the well known exponential drop of \Jc\ with misorientation is
modified and damped. Approaching the intragranular current density, coated conductors with a grain
alignment as large as 10\degree and grain aspect ratios of 50 have the same critical current
density as standard coated conductors (aspect ratio 1) with a misalignment of only 2\degree.
Networks with \ARs\ $\sim100$ support critical current densities exceeding \tenAcm{6} for grain
alignment spreads as large as $\sigma=$\DEG{25}.

Although the presented calculations consider particularly simple, mathematically accessible tape
structures and neglect self field and second order effects, they clearly show the usefulness of
optimizing the grain structure in the coated conductor technologies, in particular the use of
grains with large \ARs.

The realization of coated conductors with big aspect ratios we consider to be technologically
straightforward. The IBAD or ISD processes may be modified so that grains with large aspect ratios
are nucleated, for example, by taking advantage of anisotropic diffusion during grain nucleation
and growth. The RABiTS architecture is particularly suited for the implementation of grains with
big aspect ratios. For example, metallic tapes with standard RABiTS texture, consisting of
Ni-alloys or steel, may be rolled and annealed in mass production processes to contain long grains
which are aligned parallel to the length of the tape. Standard RABiTS buffer layer and
superconductor epitaxy, performed  by cheap deposition processes under development, will reproduce
this grain structure in the ReBa$_2$Cu$_3$O$_{7-\delta}$\xspace-based superconductor, yielding at
competitive costs high-\Tc tapes with very large critical currents.

In summary, suggesting a solution to the grain boundary problem, we propose a practical road to
competitive high-\Tc cables: fabricating doped coated conductors containing grains with big aspect
ratios.


\vspace*{0.25cm} \baselineskip=10pt{\small \noindent Interactions with U. Miller at the beginning
of the project are gratefully acknowledged. This work was financially supported by the
Bundesministerium f\"{u}r Forschung und Technologie (project 13N6918).}


\end{document}